\documentclass[amssymb,aps,prd]{revtex4}
\usepackage{epsfig}

\begin{document}



\title{{\large {\bf High Density Effective Theory of QCD}}}
\author{Deog Ki Hong\thanks{dkhong@pnu.edu} \\
Department of Physics\\
Pusan National University, Pusan 609-735, Korea \\
}
\date{\today}
\begin{abstract}
We discuss the salient features of the high density effective theory (HDET) of QCD,
elaborating more on the matching for vector-vector correlators
and axial-vector-vector correlators, which are related to screening mass and
axial anomaly, respectively.
We then apply HDET to discuss various color-superconducting phases of dense QCD.
We also review a recent proposal to solve the sign problem in dense fermionic
matter, using the positivity property of HDET.
Positivity of HDET allows us to establish rigorous inequalities in QCD
at asymptotic density and to show vector symmetry except the fermion
number is not spontaneously broken at asymptotic density.
\end{abstract}

\maketitle

\section{Introduction}

As physics advances, its frontier has expanded. One of the frontiers
under active exploration is matter at extreme conditions.
Recent surprising data, obtained from heavy-ion collisions
and compact stars such as neutron stars, and also some theoretical breakthroughs
have stimulated active investigation in this field~\cite{Rajagopal:2000wf}.

How does matter behave as we squeeze it extremely hard? This question is
directly related to one of the fundamental questions in Nature; what are the fundamental
building blocks of matter and how they interact.
According to QCD,  matter at high density is quark matter, since
quarks interact weaker  and weaker as they are put closer and closer.

At what temperature and density does the phase transition to quark matter
occur?
To determine the phase diagram of thermodynamic QCD is an outstanding problem.
The phases of matter are being mapped out by colliding heavy-ions
and by observing compact stars.
Since QCD has only one intrinsic scale,
$\Lambda_{\rm QCD}$, the phase transition of QCD matter should occur at that scale
as matter is heated up or squeezed down.
Indeed, recent lattice QCD calculations found the phase transition
does occur at temperature around $175~{\rm MeV}$~\cite{Karsch:2000kv}.
Even though lattice QCD has been quite successful at finite temperature but
at zero density, it has not made much progress at finite density
due to the notorious sign problem. The lattice calculation is usually done
in Euclidean space and Euclidean QCD with a chemical potential has a complex
measure, which precludes use of importance samplings, the main technique in
the Monte Carlo simulation for lattice calculations.

Lattice QCD at finite density is described by a partition function
\begin{eqnarray}
Z(\mu)=\int {\rm d}A\det \left(M\right)e^{-S(A)}\,,
\label{qcd_partition}
\end{eqnarray}
where $M=\gamma_E^{\mu}D_E^{\mu}+\mu\gamma_E^4$ is the Dirac operator of
Euclidean QCD with a chemical potential $\mu$. The eigenvalues of $M$
are in general complex, since $\gamma_E^{\mu}D_E^{\mu}$ is anti-Hermitian
while $\mu\gamma_E^4$ is Hermitian.  For certain gauge fields such as
$A^{\mu}(-x)=-A^{\mu}(x)$, $M$ can be mapped into $M^{\dagger}$ by a similarity
transformation and thus its determinant $M$ is nonnegative.
However,  for generic fields $M\ne P^{-1}M^{\dagger}P$ and
$\det \left(M\right)$ is complex.

Recently there have been some progress in lattice simulation
at small chemical potential, using a re-weighting method, to find the phase
line~\cite{Fodor:2001au,Allton:2002zi}.
Another interesting progress in lattice simulation was made
at very high density in~\cite{Hong:2002nn,Hong:2003zq}, where it was
shown that for QCD at high density the sign problem is either mild or absent,
since the modes, responsible for the complexness of the Dirac determinant,
decouple from dynamics or become irrelevant at high baryon density.

\section{High Density Effective Theory }

At low temperature or energy, most degrees of freedom
of quark matter are irrelevant due to Pauli blocking.
Only quasi-quarks near the Fermi surface
are excited. Therefore, relevant modes for quark matter are
quasi-quarks near the Fermi surface and the physical properties of quark matter like
the symmetry of the ground state are determined by those modes.
High density effective theory (HDET)~\cite{Hong:1998tn,Hong:1999ru}
of QCD is an effective theory for such modes
to describe the low-energy dynamics of quark matter.

To find out the modes near the Fermi surface, one needs to know the energy
spectrum of QCD, which is very difficult in general since it is equivalent
to solving QCD. However, at high density $\mu\gg\Lambda_{\rm QCD}$,
quarks near the Fermi surface carry large momenta and the typical interaction
involves a large momentum transfer. Therefore,
due to the asymptotic freedom of QCD, the spectrum near the Fermi surface
at high density looks very much like that of free fermion:
$\left(\vec\alpha\cdot \vec p-\mu+\beta\, m\right)\psi_{\pm}=E_{\pm}\psi_{\pm}\,$,
as shown in Fig.~1.
 \begin{figure}
      \centerline{\includegraphics[scale=0.4]{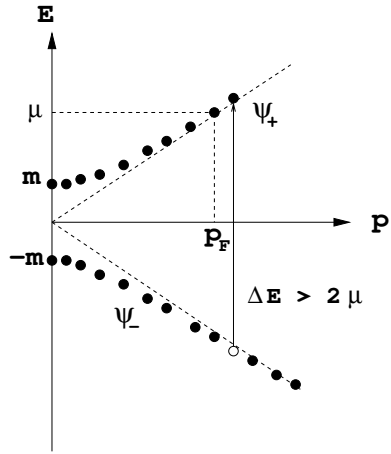}}
   \caption{Energy spectrum of quarks at high density}
   \label{fig:1}
   \end{figure}
We see that at low energy, $E<2\mu$, the states near the Fermi surface
($|\vec p|\simeq p_F$), denoted as $\psi_+$,
are easily excited while states deep in the Dirac sea, denoted as $\psi_-$, are
hard to excite.

At low energy, the typical momentum transfer by quarks near the Fermi surface
is much smaller than the Fermi momentum. Therefore, similarly to the heavy quark
effective theory, we may decompose the momentum of quarks near the Fermi surface as
\begin{equation}
p^{\mu}=\mu\,v^{\mu}+l^{\mu},
\end{equation}
where $v^{\mu}=(0,\vec v_F)$ and $\vec v_F$ is the Fermi velocity.
For quark matter, the typical size of the residual momentum is
$|l^{\mu}|\sim\Lambda_{\rm QCD}$, and
the Fermi velocity of the quarks does not change
for $\mu\gg\Lambda_{\rm QCD}$, when they are scattered off by soft gluons.

We now introduce patches to cover the Fermi surface, as shown in Fig.~2.
The sizes of each patch are $2\Lambda$ in vertical direction to the Fermi surface
and $2\Lambda_{\perp}$ in horizontal direction. The quarks in a patch are treated
to carry a same Fermi velocity.
\begin{figure}
       \centerline{\includegraphics[scale=0.5]{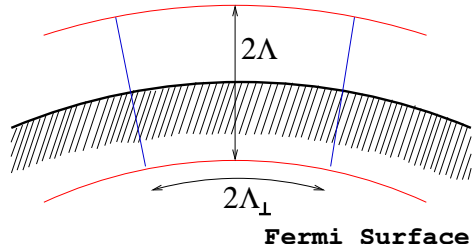}}
   \caption{A patch covering the Fermi surface}
   \label{fig:2}
   \end{figure}

The energy of the quarks in the patch is given as
\begin{equation}
E=-\mu+\sqrt{p^2+m^2}=\vec l\cdot \vec v_F+{{l}^2\over 2\mu}+O\left({1\over\mu^2}\right).
\end{equation}
We see that at the leading order in $1/\mu$ expansion,
the energy is independent of the residual momentum, $\vec l_{\perp}$,
perpendicular to the Fermi velocity. In HDET, therefore, the perpendicular
momentum labels the degeneracy and should satisfy a normalization
condition
\begin{equation}
\sum_{\rm patches}\int_{\Lambda_{\perp}} {\rm d}^2l_{\perp}=4\pi p_F^2.
\end{equation}

To identify the modes near the Fermi surface, we expand the quark
field as
\begin{equation}
\Psi(x)=\sum_{\vec v_F}e^{-i\mu \vec x\cdot\vec v_F}\left[\psi_+(\vec v_F,x)
+\psi_-(\vec v_F,x)\right],
\end{equation}
where $\psi_{\pm}(\vec v_F,x)$ satisfies respectively
\begin{equation}
{1\pm\vec \alpha\cdot\hat v_F\over2}\psi_{\pm}=\psi_{\pm}.
\end{equation}
Note that the projection operator $P_{\pm}=(1\pm\vec \alpha\cdot\hat v_F)/2$
projects out the particle state, $\psi_+$, and
the anti-particle state, $\psi_-$ (or more precisely $\bar\psi_-$),
from the Dirac spinor field $\Psi$.
The quasi-quarks in a patch carries the residual momentum $l^{\mu}$ and
is given as
\begin{eqnarray}
\psi_+(\vec v_F,x)={1+\vec \alpha\cdot\hat v_F\over2}
e^{-i\mu\vec v_F\cdot \vec x}\psi(x)
\end{eqnarray}
The Lagrangian for quark fields becomes
\begin{eqnarray}
{\cal L}\!\!&=&\!\!\bar\Psi\left(i\!\not\!\! D+\mu\gamma^0\right)\Psi
=\sum_{\vec v_F}\bar\psi\left(P_++P_-\right)
\left(\mu\!\!\!\not V\,+\,\!\!\not \!\! D\right)\left(P_++P_-\right)\psi
\nonumber\\
\!\!&=&\!\!\bar\psi_+\,i\!\not\!\! D_{\parallel}\,\psi_+
+\bar\psi_-(2\mu\gamma^0+i\!\not\!\! D_{\parallel})\psi_-
+\left[\bar\psi_-\,i\!\not\!\! D_{\perp}\,\psi_++{\rm h.c.}\right],
\end{eqnarray}
where we neglected the quark mass term for simplicity and
$V^{\mu}=(1,\vec v_F)$.
The parallel component of the covariant derivative is
$D^{\mu}_{\parallel}=V^{\mu}\,D\cdot V$ and the perpendicular component
$D_{\perp}=D-D_{\parallel}$. From the quark Lagrangian one can read off the
propagators for $\psi_{\pm}(\vec v_F,x)$:
\begin{equation}
S_F^+=P_+{i\over \not l_{\parallel}}\,,\quad S_F^-=P_-{i\gamma^0\over2\mu}
\left[1-{i\gamma^0\!\!\not l_{\parallel}\over 2\mu}+\cdots\right].
\end{equation}
We see indeed that in HDET the quarks near the Fermi surface or $\psi_+$ modes
are the propagating modes, while $\psi_-$ are not.

By integrating out $\psi_-$ modes and the hard gluons, one obtains the high
density effective theory of QCD. In general the integration results in
nonlocal terms in the effective theory and one needs to expand them
in powers of $1/\mu$. This is usually done by matching the one-light-particle
irreducible amplitudes of the microscopic theory with those of the effective theory.
For tree-level amplitudes, this is tantamount to eliminating the irrelevant modes,
using the equations of motion.
\begin{eqnarray}
\label{eliminate} \psi_-(\vec v_F,x)=-{i\gamma^0\over 2\mu
+i\!\not\!\!D_{\parallel}}{D\!\!\!\!/}_{\perp}\psi_+=-
{i\gamma^0\over
2\mu}\sum_{n=0}^{\infty}\left(-{i\!\not\!\!D_{\parallel}\over 2\mu}\right)^n
{D\!\!\!\!/}_{\perp}\psi_+\,. \label{eom}
\end{eqnarray}
For instance, a one-light particle irreducible amplitude in QCD of two
gluons and two quarks is matched as
\begin{equation}
\bar\psi_+
i\not\!\!D_{\perp}\psi_-(\vec v_F,x)
\bar\psi_-i\not\!\!D_{\perp}\psi_+(\vec v_F,y)
=\bar\psi_+
i\not\!\!D_{\perp}\left(-{i\gamma^0\over 2\mu}\right)
i\not\!\!D_{\perp}\psi_+\,,
\end{equation}
which is shown in Fig.~3.
\begin{figure}
       \centerline{\includegraphics[scale=0.5]{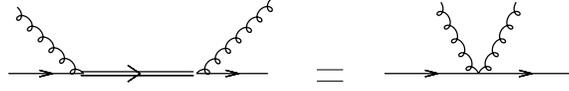}}
   \caption{Tree-level matching: The double line denotes $\psi_-$ modes and
   the single line  $\psi_+$.}
   \label{fig:3}
   \end{figure}
Similarly one can eliminate the hard gluons.
Integrating out hard gluons results in four-Fermi interactions of $\psi_+$ modes.
(See Fig.~4.)
\begin{figure}
       \centerline{\includegraphics[scale=0.5]{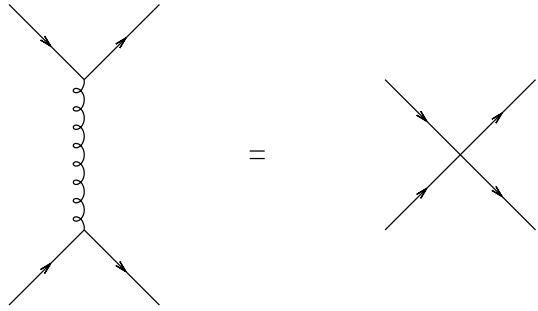}}
   \caption{Tree-level matching: Four-Fermi interaction due to hard gluons}
   \label{fig:4}
   \end{figure}
One continues matching one-loop or higher-loop amplitudes.
One interesting feature of HDET is that a new marginal operator arises at
the one-loop matching, when incoming quarks are in Cooper-paring kinematics,
namely when they have opposite Fermi velocities.
As shown in Fig.~5, when the incoming quarks have opposite Fermi velocity,
the amplitudes in HDET are ultra-violet divergent while QCD amplitudes are not.
Therefore, one needs to introduce a four-Fermi operator as a counter term
to remove the UV divergence.
\begin{figure}
       \centerline{\includegraphics[scale=0.7]{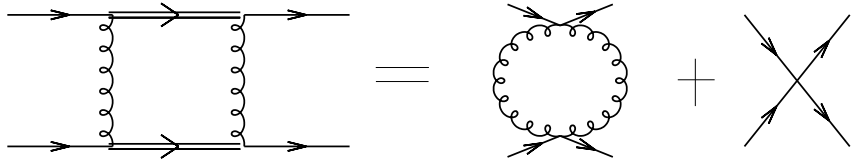}}
   \caption{One-loop matching}
   \label{fig:5}
   \end{figure}
If we collect all the terms in the effective theory, it has a systematic expansion
in $1/\mu$ and coupling constants $\alpha_s$ as
\begin{equation}
\label{treeL} {\cal L}_{\rm HDET}=
b_1\bar\psi_+i\gamma_{\parallel}^{\mu}D_{\mu}\psi_+-{c_1\over2\mu}\bar\psi_+
\gamma^0({D\!\!\!\!/}_{\perp})^2\psi_+ ~+~ \cdots,
\end{equation}
where $b_1=1+O(\alpha_s),~c_1=1+O(\alpha_s),\cdots$. Note that HDET has
a reparametrization
invariance, similarly to heavy quark effective theory, which is due to the fact
that the Fermi velocity of quarks in a patch is not uniquely determined.
For a given quark momentum, the corresponding Fermi
velocity is determined up to reparametrization; $\vec v_F\to \vec
v_F+\delta \vec l_{\perp}/\mu$ and $\vec l\to \vec l-\delta \vec l$,
where $\delta \vec l_{\perp}$ is a residual momentum perpendicular
to the Fermi velocity.
As in the heavy quark effective theory~\cite{manohar92},
the renormalization of higher-order operators are restricted due to the
reparametrization invariance. For instance, $b_1=c_1$ at all orders in $\alpha_s$.

In order for the effective theory to be meaningful, it should have
a consistent power-counting. We find the consistent counting in HDET to be
for $\Lambda_{\perp}=\Lambda$
\begin{equation}
\left({D_{\parallel}\over\mu}\right)^n\cdot\left({D_{\perp}\over\mu}\right)^m
\cdot\psi^l_+\sim \left({\Lambda\over\mu}\right)^{n+m}\Lambda^{3l/2}.
\end{equation}
To be consistent with the power counting, we impose in loop integration
\begin{equation}
\int_{\Lambda_{\perp}}{\rm d}^2l_{\perp}\,l_{\perp}^n=0\quad {\rm for}\quad
n>0.
\end{equation}
So far we have restricted ourselves to operators containing quarks in
the same patch. For operators with quarks in different patches, one
has to be careful, since the loop integration might jeopardize the
power-counting rules. Indeed,  consistent counting is to sum up all the hard-loops,
as shown by Sch\"afer~\cite{Schafer:2003jn}.

\section{More on matching}
In HDET, the currents are given in terms of particles
and holes but without antiparticles as
\begin{eqnarray}
J^{\mu}=\sum_{\vec v_F}\bar\psi(\vec v_F,x)\gamma^{\mu}_{\parallel}
\psi(\vec v_F,x)-{1\over 2\mu}\psi^{\dagger}
\left[\gamma^{\mu}_{\perp},\,i\!\not\!\!D_{\perp}\right]
\psi+\cdots\,,
\end{eqnarray}
where the color indices are suppressed and
we have reverted the notation $\psi$ for $\psi_+$ henceforth.
We find that the HDET current is not conserved
unless one adds a counter term. Consider the current correlator
\begin{eqnarray}
\left<J^{\mu}(x)J^{\nu}(y)\right>&=&{\delta^2 \Gamma_{\rm eff}\over
\delta A_{\mu}(x)\delta A_{\nu}(y)}=\int_pe^{-ip\cdot(x-y)}\Pi^{\mu\nu}(p)
\end{eqnarray}
where the vacuum polarization tensor
\begin{eqnarray}
\Pi^{\mu\nu}_{ab\rm}(p)
&=&-{iM^2\over2}\delta_{ab} \int{{\rm d}\Omega_{\vec v_F}\over 4\pi}
\left({-2\vec p\cdot\vec v_FV^{\mu}V^{\nu}\over p\cdot V
+i\epsilon \vec p\cdot\vec v_F}\right)
\end{eqnarray}
and  $M^2=N_f g_s^2\mu^2/(2\pi^2)$.
We see that the vacuum polarization tensor is not transverse,
$p_{\mu}\Pi^{\mu\nu}_{ab}(p)\ne0$,
which means that the current is not conserved. The physical reason for this
is that not only modes near the Fermi surface but also modes deep in the
Fermi sea respond to external sources collectively.
To recover the current conservation in the effective theory,
we need to add the DeBye screening mass term due to $\psi_-$ (See Fig.~6):
\begin{equation}
\Gamma^{\rm eff}\mapsto \tilde\Gamma^{\rm eff}=\Gamma^{\rm eff}-\int_x{M^2\over2}
\sum_{\vec v_F}A_{\mu}A_{\nu}g^{\mu\nu}_{\perp}.
\end{equation}
Then vacuum polarization tensor becomes
\begin{equation}
\Pi^{\mu\nu}(p)\mapsto\tilde\Pi^{\mu\nu}(p)
=\Pi^{\mu\nu}-{i\over2}\sum_{\vec v_F}g^{\mu\nu}_{\perp}M^2\,.
\end{equation}
The modified polarization tensor is now transverse, $p_{\mu}\tilde\Pi^{\mu\nu}=0$.
\begin{figure}
       \centerline{\includegraphics[scale=0.5]{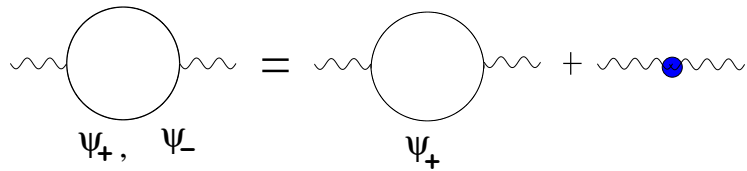}}
   \caption{Matching two-point functions}
   \label{fig:5}
   \end{figure}

Now, let us consider the divergence of axial currents in HDET, which is related
to the axial anomaly and also to how the quark matter responds to external
axial-current sources like electroweak probes.

It is easy to show that the axial anomaly in dense matter is independent
of density or the chemical potential $\mu$~\cite{hong}.
In general one may re-write the divergence of axial currents in dense QCD
as follows:
\begin{equation}
\left<\partial_{\mu}J^{\mu}_5\right>={g^2\over16\pi^2}
{\tilde F}_{\mu\alpha}F^{\mu\alpha}+
\Delta^{\alpha\beta}(\mu)A_{\alpha}A_{\beta}\,,
\end{equation}
where the first term is the usual axial anomaly in vacuum and the second
term is due to matter. However, one can explicitly calculate the  second
term, which is finite, to find $ \Delta^{\alpha\beta}(\mu)=0$.
In HDET, the axial anomaly due to modes near the Fermi surface is
given as
\begin{eqnarray}
\sum_{\vec v_F}\int_{x,y}e^{ik_1\cdot x+ik_2\cdot y}
\left<\partial_{\mu}J^{\mu}_5(\vec v_F,0)
J^{\alpha}(\vec v_F,x)J^{\beta}(\vec v_F,y)\right>
\equiv\Delta^{\alpha\beta}_{\rm eff}
\end{eqnarray}
By explicit calculation we find
\begin{eqnarray}
\Delta^{0i}_{\rm eff}=-{g^2\over4\pi^2}
\cdot{{1\over3}}\left(\vec k_1\times\vec k_2\right)^i,
~\Delta^{ij}_{\rm eff}={g^2\over4\pi^2}{{2\over3}}
\epsilon^{ijl}\left(k_{10}k_{2l}-k_{1l}k_{20}\right)\,.
\end{eqnarray}
We see that the modes near the Fermi surface contributes only some parts
of the axial anomaly. As in the vector current, the rest should come
from modes in the deep Fermi sea and from anti-particles. To recover the full
axial anomaly we add a counter term (See Fig.~7), which is two thirds of the
axial anomaly plus a Chern-Simons term:
\begin{equation}
\tilde\Delta^{\alpha\beta}_{\rm eff}=\Delta^{\alpha\beta}_{\rm eff}
+{g^2\over6\pi^2}\epsilon^{\alpha\beta\rho\sigma}k_{1\rho}k_{2\sigma}
+{g^2\over12\pi^2}\epsilon^{\alpha\beta 0l}\left(
k_{10}k_{2l}-k_{1l}k_{20}\right).
\end{equation}
\begin{figure}
       \centerline{\includegraphics[scale=0.5]{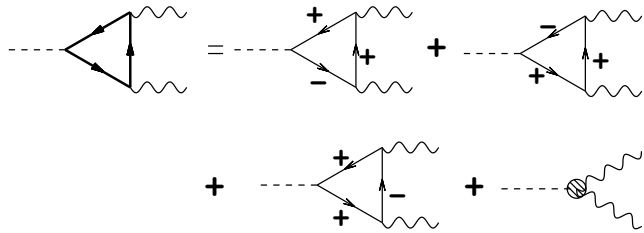}}
   \caption{Matching axial anomaly. $+$ denotes $\psi_+$ and $-$
   denotes $\psi_-$.}
   \label{fig:7}
   \end{figure}

\section{Color superconductivity in dense QCD}
At high density, quarks in dense matter interact weakly with
each other and form a Fermi sea, due to asymptotic freedom.
When the energy is much less than the quark chemical potential
($E\ll\mu$), only the quarks near the Fermi surface are relevant.
The dynamics of quarks near the Fermi surface
is effectively one-dimensional, since excitations along the Fermi
surface do not cost any energy. The momentum perpendicular to the
Fermi momentum just labels the degeneracy, similarly to the
perpendicular momentum of charged particle under external
magnetic field. This dimensional reduction due to the presence of
Fermi surface makes possible for quarks to form a Cooper pair
for any arbitrary weak attraction, since the critical coupling
for the condensation in (1+1) dimensions is zero, known
as the Cooper theorem in condensed matter.

While, in the BCS theory, such attractive force for electron Cooper pair
is provided by phonons,
for dense quark matter, where phonons are absent,
the gluon exchange interaction provides the attraction, as one-gluon
exchange interaction is attractive in the color anti-triplet
channel\footnote{There is also an attractive force between quarks and holes in
the color octet channel: $\left<\bar\psi_i(-\vec p)\psi_j(\vec p)\right>\ne0$,
which corresponds to a density wave. However, because of the momentum conservation,
the density wave condensate does not enjoy the full Fermi surface degeneracy. Indeed,
for QCD, the diquark condensate is energetically preferred to the density
wave condensate~\cite{Shuster:1999tn}.}
One therefore expects that color anti-triplet Cooper pairs will
form and quark matter is color
superconducting, which is indeed shown more than 20 years
ago~\cite{Bar_77,Frau_78,BL_84}.

At intermediate density, quarks and gluons are strongly interacting and
gluons are therefore presumably screened. Then, QCD at intermediate density
may be modelled by four-Fermi interactions and higher-order terms by massive gluons.
\begin{equation}
{\cal L}^{\rm eff}_{\rm QCD}\ni {G\over 2}
\bar\psi\psi\bar\psi\psi+\cdots,
\end{equation}
where the ellipsis denotes higher-order terms induced by massive gluons.
When the incoming quarks have opposite momenta, the four-Fermi interaction
is marginally relevant, if attractive, and all others are irrelevant.
As the renormalization group flows toward
the Fermi surface, the attractive four-Fermi interaction is dominant and blows up,
resulting in a Landau pole, which can be avoided only when a gap opens at the Fermi
surface. This is precisely the Cooper-instability of the Fermi surface.
The size of gap can be calculated by solving the gap equation, which is
derived by the variational principle that the gap minimizes the vacuum energy:
\begin{eqnarray}
0={\partial V_{\rm BCS} (\Delta)\over\partial \Delta}=
{\Delta \over G}- i\int{d^4k\over (2\pi)^4}{\Delta\over
k_0^2-({\vec k\cdot\vec v_F})^2-\Delta^2},
\end{eqnarray}
which gives
\begin{equation}
\Delta=-iG\,\int{d^4k\over (2\pi)^4}{\Delta\over
\left[(1+i\epsilon)k_0\right]^2-({\vec k\cdot\vec v_F})^2-\Delta^2}.
\label{gap}
\end{equation}
We note that the integrand in Eq.~\ref{gap} does not depend on $k_{\perp}$,
whose integration gives the density of states at the Fermi surface, and the
$i\epsilon $ prescription is consistent with the Feynman propagator.
The pole occurs at
\begin{equation}
k_0=\pm \sqrt{({\vec k\cdot\vec v_F})^2+\Delta^2}\,\,\mp i\epsilon
\end{equation}
or in terms of full momentum $p=\mu\, v+k$ it occurs at
\begin{equation}
p_0=\pm \sqrt{(|\vec p|-\mu)^2+\Delta^2}\,\,\mp i\epsilon\,.
\end{equation}
We find the solution to the gap equation
\begin{equation}
\Delta_0=2\mu\exp\left(-{\pi^2\over 2G{\mu}^2}\right).
\end{equation}
For generic parameters of dense QCD, the gap is estimated to be
$10\sim100\,{\rm MeV}$ at the intermediate density.
The free energy of the BCS state is given as
\begin{eqnarray}
V_{\rm BCS}(\Delta_0)&=&\int_0^{\Delta_0}{\partial V_{\rm BCS}\over\partial\Delta}
{\rm d}\Delta\nonumber\\
&=&{4\mu^2\over G}\int_0^{x_0}\left(x+g^2\ln x\right)dx=-
{\mu^2\over 4\pi^2}\Delta_0^2,
\end{eqnarray}
where $x=\Delta/(2\mu)$ and $g^2=2G\mu^2/\pi^2$.
At high density magnetic gluons are not screened though electric gluons are
screened~\cite{Baym:im,Pisarski:1998nh,Son:1998uk}.
The long-range pairing force mediated by magnetic gluons leads to the Eliashberg
gap equation (See Fig.~8).
\begin{eqnarray}
\Delta(p_0)&=&{g_s^2\over 36\pi^2}\int_{-\mu}^{\mu}dq_0
{\Delta(q_0)\over \sqrt{q_0^2+\Delta^2}}
\ln\left({{\bar\Lambda}\over |p_0-q_0|}
\right),
\label{gapf}
\end{eqnarray}
where $\bar\Lambda=4\mu/\pi\cdot (\mu/M)^5e^{3/2\xi}$ and
$\xi$ is a gauge parameter.
\begin{figure}
\vskip 0.2in
       \centerline{\includegraphics[scale=0.6]{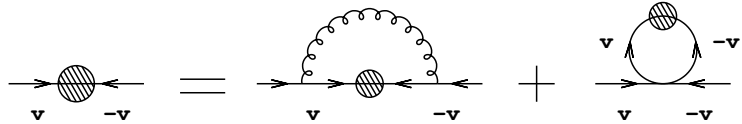}}
   \caption{Eliashberg equation at high density.}
   \label{fig:8}
   \end{figure}
Due to the unscreened but Landau-damped gluons, there is an extra (infrared)
logarithmic divergence in the gap equation, when the incoming quark momentum
is collinear with the gluon momentum.
The Cooper-pair gap at high density is found to be~\cite{Son:1998uk,Hong:1999fh}
\begin{equation}
\Delta_0=
{2^7\pi^4\over N_f^{5/2}}e^{3\xi/2+1}\cdot{\mu\over g_s^5}
\exp\left(-{3\pi^2\over\sqrt{2}g_s}\right).
\nonumber
\end{equation}
The numerical prefactor of the gap is not complete, since the contributions
from subleading corrections to the gap equation
that lead to logarithmic divergences, such as
the wavefunction renormalization  and the vertex corrections,
are not taken into account. Recently, however, the contributions to the prefactor,
coming from the vertex corrections and the wavefunction renormalization
for quarks were calculated by finding  a (nonlocal) gauge~\cite{Hong:2003ts},
where the quark wavefunction is not renormalized for all momenta, $Z(p)=1$.
At the nonlocal gauge, $\xi\simeq1/3$. The subleading corrections therefore increase
the leading-order gap at the Coulomb gauge by about two thirds.

\section{Quark matter under stress}
It is quite likely to find dense quark matter inside compact stars like
neutron stars. However, when we study the quark matter in compact stars,
we need to take into account not only the charge and color neutrality
of compact stars and but also the mass of the strange quark,
which is not negligible at the intermediate density.
By the neutrality condition and the strange quark mass,
the quarks with different quantum numbers in general have different
chemical potentials and different Fermi momenta. When the
difference in the chemical potential becomes too large the Cooper-pairs breaks
or other exotic phases like kaon condensation or crystalline phase
is more preferred to the BCS phase.

Let us consider for example the pairing between up and strange quarks
in chemical equilibrium. The energy spectrum of up quarks is given as
\begin{equation}
E=-\mu \pm |\vec p|,
\end{equation}
while the energy of strange quarks of mass $M_s$ becomes
\begin{equation}
E=-\mu\pm\sqrt{|\vec p|^2+M_s^2}\,.
\end{equation}
The Fermi sea of up and strange quarks is shown in Fig.~9. Because of the strange
quarks mass, they have different Fermi momenta.
\begin{figure}
       \centerline{\includegraphics[scale=0.4]{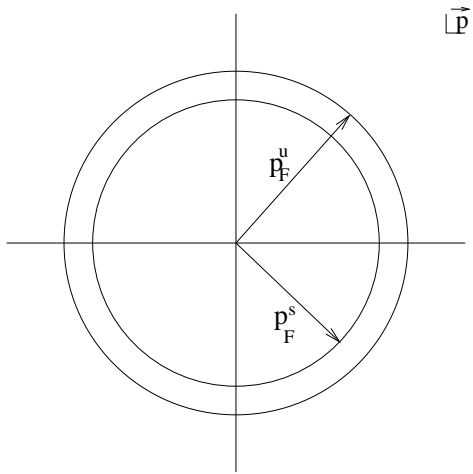}}
   \caption{Fermi sea of up and strange quarks.}
   \label{fig:9}
   \end{figure}
Note that the Cooper-pairing occurs for quarks with same but opposite momenta.
Therefore, at least one of the pairing quarks should be excited away from
the Fermi surface, costing some energy. Let us suppose that the Cooper-pair
gap opens at $|\vec p|=\bar p$ between two Fermi surfaces,
$p_F^s\le \bar p\le p_F^u$.

To describe such pairing, we consider small fluctuations of up and strange quarks
near $\bar p$. The energy of such fluctuations of up and down quarks is respectively
\begin{equation}
E_u=-\mu+|\bar p+\vec l\,|\simeq -\delta\mu^{u}+\vec v_u\cdot \vec l,
\quad\quad E_s\simeq-\delta\mu^s+\vec v_s\cdot\vec l,
\end{equation}
where $\delta\mu^u=\mu-\bar p$ and $\delta\mu^s=\mu-\sqrt{{\bar p}^2+M_s^2}$.
$\vec v_u$ and $\vec v_s$ are the velocities of up and strange quarks
at $|\vec p|=\bar p$.
Let $\Delta$ be the BCS gap for the $u,\,s$ pairing. Then, the Lagrangian for the
$u,\,s$ quarks is given as
\begin{eqnarray}
{\cal L}=\bar u \left(i\not\!\partial+\mu\gamma^0\right)u+
\bar s_c\left(i\not\!\partial-\!\mu\gamma^0\!-\!M_s\right)s_c\!-\!
\Delta\,\bar s_c\,u+{\rm h.c.}+{\cal L}_{\rm int},
\end{eqnarray}
where $s_c$ is the charge conjugate field of $s$ quark.
In HDET, the Lagrangian becomes
\begin{eqnarray}
{\cal L}_{\rm HDET}\ni u^{\dagger}\left(iV_u\cdot \partial +\delta\mu^u\right)u
+s_c^{\dagger}\left(i\bar V_s\cdot \partial-\delta\mu^s\right)s_c
-\Delta \bar s_c\,u+{\rm h.c.},
\end{eqnarray}
where $V_u=(1,\vec v_u)$ and $\bar V_s=(1, -\vec v_s)$.
The Cooper-pair gap equation is then
\begin{equation}
\Delta(p)\!=\!\int_l
{i\Delta(l)\,K(p-l)\over \left[(1+\!i\epsilon)l_0-\vec l\cdot \vec v_u+\!\delta\mu^u\right]
\left[(1+\!i\epsilon)l_0+\vec l\cdot \vec v_s-\!\delta\mu^u\right]\!-\!\Delta^2}\,,
\end{equation}
where $K$ is the kernel for the gap equation and is a constant for
the four-Fermi interaction.
By examining the pole structure, we see that the Cooper-pair gap does not exist when
\begin{equation}
-\delta\mu^u\delta\mu^s>{\Delta^2\over4}.
\end{equation}
Only when $-\delta\mu^u\delta\mu^s<\Delta^2/4$, one can shift
$l_0\to l_0^{\prime}=l_0+\delta\mu^u$ or $l_0\to l_0^{\prime}=l_0-\delta\mu^s$
without altering the pole structure.
Note that the gap becomes biggest when $\delta\mu^u=-\delta\mu^s(\equiv\delta\mu)$,
which determines the pairing momentum to be
\begin{equation}
\bar p=\mu-{M_s^2\over4\mu}.
\end{equation}
If $\delta\mu<\Delta/2$ or $\Delta>M_s^2/(2\mu)$, the solution to the Cooper-pair
gap exists. The gap equation then can be written as, shifting $l_0$, in Euclidean
space
\begin{equation}
\Delta(p)=\int{{\rm d}^4l\over (2\pi)^4}\,
{\Delta(l)\over l_{\parallel}^2+\Delta^2}\,K(l-p),
\end{equation}
where
$l_{\parallel}^2=l_0^2+c^2(\vec l\cdot \hat v)^2$ and
$c^2=\bar p/\sqrt{{\bar p}^2+M_s^2}\,$.
In HDET, one can easily see that the
Cooper-pair gap closes if the effective chemical potential difference,
2$\delta\mu$, due to an external stress, exceeds the Cooper-pair
gap when there is no stress. One should note that even before the Cooper-pair
gap closes other gap may open as shown by many
authors~\cite{Alford:2000ze,Bedaque:2001je}. But, one needs to
compare the free energy of each phases to find the true ground state
for quark matter under stress.

\section{Positivity of HDET}

Fermionic dense matter generically suffers from the sign problem,
which has thus far precluded lattice
simulations~\cite{Hands:2001jn}.
However, the sign problem usually associated
with fermions is absent if one considers only low-energy degrees
of freedom.
The complexness of the measure of
fermionic dense matter can be ascribed to modes far from the Fermi
surface, which are irrelevant to dynamics at sufficiently high density
in most cases, including quark matter~\cite{Hong:2002nn,Hong:2003zq}.
For modes near the Fermi surface,
there is a discrete symmetry, relating particles and holes,
which pairs the eigenvalues of the Dirac operator to make its
determinant real and nonnegative.
Especially, the low energy effective theory of dense QCD has positive
Euclidean path integral measure, which allows one to establish
rigorous inequalities that the color-flavor locked (CFL) phase
is the true vacuum of three flavor, massless QCD.

As simple example, let us consider a fermionic matter in 1+1 dimensions,
where non-relativistic fermions are interacting with a gauge field $A$.
The action is in general given as
\begin{eqnarray}
\label{NRA} S\!\!=\!\! \int\!\! d\tau dx~ \psi^{\dagger} \left[ ( -
\partial_\tau + i\phi + \epsilon_F ) - \epsilon( -i\partial_x +
A ) \right] \psi,
\end{eqnarray}
where $\epsilon(p)\simeq p^2/(2m)+\cdots$ is the energy as a function of
momentum.
Low energy modes have momentum near the Fermi points and have energy,
measured from the Fermi points,
\begin{equation}
E(p\pm p_F)\simeq \pm\, v_F p, \quad v_F=\left.{\partial E\over \partial p}\right\vert_{p_F}.
\end{equation}
If the gauge fields have small amplitude and are slowly varying relative to
scale $p_F$, the fast modes are decoupled from low energy physics. The low energy
effective theory involving quasi particles and gauge fields has a positive,
semi-definite determinant.

To construct the low energy effective theory of the fermionic system, we rewrite
the fermion fields as
\begin{equation}
\psi (x, \tau) =\psi_L(x,\tau) e^{+i p_F x} ~+~\psi_R(x,\tau) e^{-i p_F x},
\end{equation}
where $\psi_{L,R}$ describes the small fluctuations of quasiparticles near the
Fermi points.
Using $
e^{\pm ip_F x} ~E( -i\partial_x + A )~ e^{\mp ip_F x} ~\psi (x)
\approx \pm~ v_F (-i\partial_x + A) \psi(x),$
we obtain
\begin{eqnarray}
\label{EFA} S_{\rm eff}\!\! =\!\! \int_{\tau,x}\!\!\left[
\psi^{\dagger}_L
(-\partial_\tau + i\phi + i\partial_x -\! A ) \psi_L
+\psi^{\dagger}_R ( - \partial_\tau + i\phi - i\partial_x +\! A ) \psi_R
\right].
\end{eqnarray}
Introducing the Euclidean (1+1) gamma matrices
$\gamma_{0,1,2}$ and
$\psi_{L,R} = \frac{1}{2} ( 1 \pm \gamma_2) \psi$, we obtain
a positive action:
\begin{equation}
\label{SReff} S_{\rm eff} = \int {d}\tau\,dx\,   \bar{\psi}
\gamma^{\mu} (\partial_{\mu} + iA_{\mu} ) \psi \equiv\!\int d\!\tau
dx\, \bar{\psi}\, D\!\!\!\!/ \,\psi .
\end{equation}
Since $D\!\!\!\!/ = \gamma_2{D\!\!\!\!/}^{\dagger}\gamma_2$,
the determinant of $D\!\!\!\!/$ is positive, semi-definite.

In this example, we see that near the Fermi surface, modes
have low energy, slowly varying, and thus
lead to an effective theory without any sign problem what so ever if
they couple to slowly varying background fields.
QCD at high baryon density falls into this category,
since the coupling constant is small at high energy
due to asymptotic freedom.

HEDT of quark matter is described by
\begin{equation}
\label{treeL} {\cal L}_{\rm HDET}=
\bar\psi_+i\gamma_{\parallel}^{\mu}D_{\mu}\psi_+-{1\over2\mu}\bar\psi_+
\gamma^0({D\!\!\!\!/}_{\perp})^2\psi_+ ~+~ \cdots,
\end{equation}
where $\gamma^{\mu}_{\parallel}=(\gamma^0,\vec v_F\vec v_F\cdot
\vec\gamma)=\gamma^{\mu}-\gamma^{\mu}_{\perp}$.
We see that the leading term has a positive determinant, since
\begin{equation}
M_{\rm eft}=~\gamma^{E}_{\parallel}\cdot D(A)~=~
\gamma_5M_{\rm eft}^{\dagger}\gamma_5.
\end{equation}

In order to implement this HDET on lattice, it is convenient to introduce
an operator formalism, where the velocity is realized as an operator,
\begin{equation}
\label{velo} \vec{v} =   \frac{-i }{\sqrt{- \nabla^2}}
~\frac{\partial}{\partial \vec{x}}~~.
\end{equation}
Then the quasi quarks near the Fermi surface become
\begin{equation}
\psi = \exp \left( + i \mu x \cdot v \right)
{1+ \alpha \cdot v\over 2}
\psi_+  .
\end{equation}
Now, neglecting the higher order terms,
the Lagrangian becomes with $X=\exp(i\mu x\cdot v)
(1+\alpha\cdot v)/2$,
\begin{eqnarray}
\label{leading2}
{\cal L}_{\rm HDET} =  \bar{\psi}_+  \gamma^\mu_\parallel
\left(
\partial^\mu + i A^\mu_+ \right) \psi_+,
\end{eqnarray}
where $A^\mu_+  =  X^{\dagger}\,A^\mu\,X$ denotes soft gluons whose
momentum $|p_{\mu}|<\mu$.
Since $v \cdot \partial \, v \cdot \gamma =
\partial \cdot \gamma~$, we get
\begin{eqnarray}
\gamma^\mu_{\parallel} \partial^\mu = \gamma^\mu \partial^\mu
\end{eqnarray}
which shows that the operator formalism automatically covers modes near
the full Fermi surface.

Integrating out the fast modes, modes far from the Fermi surface and
hard gluons, the QCD partition function~(\ref{qcd_partition}) becomes
\begin{equation}
Z(\mu)=\int {\rm d}A_+\det \left(M_{\rm eff}\right)e^{-S_{\rm eff}(A_+)},
\label{epartition}
\end{equation}
where
\begin{eqnarray}
S_{\rm eff}=\int_{x_E}\left({1\over4}F_{\mu\nu}^aF_{\mu\nu}^a +{M^2\over
16\pi}A_{\perp\mu}^{a}A_{\perp\mu}^{a}\right)
+\cdots
\end{eqnarray}
and $A_{\perp}=A-A_{\parallel}$, the Debye mass
$M=\sqrt{N_f/(2\pi^2)}g_s\mu$\,.
At high density the higher order terms ($\sim \Lambda/\mu$) are negligible
and the effective action becomes positive, semi-definite. Therefore,
though it has non-local operators, HDET in the operator formalism,
free from the sign problem, can be used to simulate the Fermi surface
physics like superconductivity. Furthermore,
being exactly positive at asymptotic density, HDET allows to establish
rigorous inequalities relating bound state masses and forbidding the
breaking of vector symmetries, except baryon number,
in dense QCD~\cite{Hong:2003zq}.

With the help of previous two examples, we propose a new way of
simulating dense QCD, which evades the sign problem.
Integrating out quarks far from the Fermi surface,
which are suppressed by $1/\mu$ at high density,
we can expand the determinant of Dirac operator at finite density,
\begin{equation}
\det \left(M\right) = \left[ {\rm real, positive} \right] \left[
1 ~+~ {\cal O} \left( \frac{ {{\bf F}}  }{\mu^2} \right) \right].
\end{equation}
As long as the gauge fields are slowly varying, compared to
the chemical potential $\mu$, the sign problem can be evaded.
As a solution to the sign problem,
we propose to use two lattices with different spacings, a finer lattice with
a lattice spacing  ${a_{\rm det}}\sim \mu^{-1}$ for fermions and a coarser lattice
with a lattice spacing  ${a_{\rm gauge}}\ll \mu^{-1}$ for
gauge fields and then compute the determinant on such  lattices.

\begin{figure}
       \centerline{\includegraphics[scale=0.3]{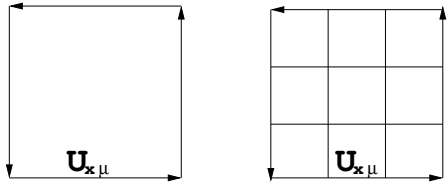}}
   \caption{Simulation with two lattices with different lattice spacings }
\label{fig}
\end{figure}

The determinant is a function of plaquettes
${\bf \{ U_{x \mu} \}}$ which are obtained by interpolation from the
plaquettes on the coarser  lattice of lattice spacing $a_{\rm gauge}$.
To get the link variables for the finer lattice, we interpolate
the link variables ${\bf U_{x \mu}} \in SU(3)$ (see Fig.~\ref{fig}):
Connect any two points $g_1, g_2$ on the group manifold as
\begin{equation}
g(t) = g_1 + t (g_2 - g_1)~,~ 0 \leq t \leq 1\,.
\end{equation}


For importance sampling in the lattice simulation,
one can use  the leading part of the
determinant, $[{\rm real}, {\rm positive}]$.
This proposal provides a nontrivial check on analytic results at
asymptotic density and can be used to extrapolate to intermediate
density. Furthermore, it can be applied  to condensed matter systems
like High-$T_c$ superconductors, which in general suffers from a
sign problem.

Positivity of the measure allows for rigorous QCD inequalities at
asymptotic density. For example, inequalities among masses of
bound states can be obtained using bounds on bare quasiparticle
propagators. One subtlety that arises is that a quark mass term
does not lead to a quasiparticle gap (the mass term just shifts
the Fermi surface). Hence, for technical reasons the proof of
non-breaking of vector symmetries~\cite{Vafa:1984xg} must be
modified. (Naive application of the Vafa-Witten theorem would
preclude the breaking of baryon number that is observed in the
color-flavor-locked (CFL) phase~\cite{Alford:1998mk}). A
quasiparticle gap can be inserted by hand to regulate the bare
propagator, but it will explicitly violate baryon number. However,
following the logic of the Vafa-Witten proof, any symmetries which
are preserved by the regulator gap cannot be broken spontaneously.
One can, for example, still conclude that isospin symmetry is
never spontaneously broken (although see below for a related
subtlety). In the case of three flavors, one can introduce a
regulator $d$ with the color and flavor structure of the CFL gap
to show rigorously that none of the symmetries of the CFL phase
are broken at asymptotic density. On the other hand, by applying
anomaly matching conditions \cite{anomaly}, we can prove that the
$SU(3)_A$ symmetries {\it are} broken. We therefore conclude that
the CFL phase is the true ground state for three light flavors at
asymptotic density, a result that was first established by
explicit calculation
\cite{Evans:1999at,Hong:1999ru,Schafer:1999fe}.

To examine the long-distance behavior of the vector current,
we note that the correlator of the vector current
for a given gauge field $A$ can be written as
\begin{eqnarray}
\left<J_{\mu}^a(\vec v_F,x)J_{\nu}^b(\vec v_F,y)\right>^A =-{\rm
Tr}\,\gamma_{\mu}T^a S^A(x,y;d)\gamma_{\nu}T^b S^A(y,x;d),
\end{eqnarray}
where the $SU(N_f)$ flavor current $J_{\mu}^a(\vec v_F,x)
=\bar\psi_+(\vec v_F,x)\gamma_{\mu}T^a\psi_+(\vec v_F,x)$. The
propagator with $SU(3)_V$-invariant IR regulator $d$ is given as
\begin{eqnarray}
S^A(x,y;d)=\left<x\right|\frac{1}{M}\left|y\right>=\int_0^{\infty}
{\rm d}\tau \left<x\right|e^{-i\tau (-iM)}\left|y\right> \nonumber
\end{eqnarray}
where with $D=\partial+iA$
\begin{eqnarray}
M &=& \gamma_0
\left(
\begin{array}{cc}
D\cdot V \quad d \\
d^{\dagger}  \quad D\cdot\bar V
\end{array}\right)
\end{eqnarray}
Since the eigenvalues of $M$ are bounded from below by $d$, we
have
\begin{eqnarray}
\left|\left<x\right|\frac{1}{M}\left|y\right>\right| \le
\int_R^{\infty}\!\!\!{\rm d}\tau \,e^{-d \,
\tau}\sqrt{\left<x|x\right>} \sqrt{\left<y|y\right>}=\frac{e^{-d
\, R}}{d} \sqrt{\left<x|x\right>}\sqrt{\left<y|y\right>},
\label{propagator1}
\end{eqnarray}
where $R\equiv\left|x-y\right|$.
The current correlators fall off rapidly as
$R\to \infty$;
\begin{eqnarray}
\left| \int {\rm d}A_+\!\!\!\right.\!\!\!\!\!& &\left.\!\!\!\!\! ~
\det M_{\rm eff}(A)\,\,e^{-S_{\rm eff}}
\left<J_{\mu}^A(\vec v_F,x)J_{\nu}^B(\vec v_F,y)\right>^{A_+}\right|\nonumber\\
\le \int_{A_+}\!\!\!& & \!\!\! \left| \left<J_{\mu}^A(\vec v_F,x)J_{\nu}^B(\vec v_F,y)
\right>^{A_+}\right| \le \frac{e^{-2d \, R}}{d^2}
\int_{A_+} \left|\left<x|x\right>\right|\left|\left<y|y\right>\right|,
\label{schwartz}
\end{eqnarray}
where we used the Schwartz inequality in the first inequality,
since the measure of the effective theory is now positive, and
equation (\ref{propagator1}) in the second inequality. The IR
regulated vector currents do not create massless modes out of the
vacuum or Fermi sea, which implies that there is no
Nambu-Goldstone mode in the $SU(3)_V$ channel. Therefore, for
three massless flavors $SU(3)_V$ has to be unbroken as in CFL. The
rigorous result provides a non-trivial check on explicit
calculations, and applies to any system in which the quasiparticle
dynamics have positive measure.

It is important to note the order of limits necessary to obtain
the above results. Because there are higher-order corrections to
the HDET, suppressed by powers of $\Lambda / \mu$, that spoil its
positivity, there may be contributions on the RHS of
(\ref{schwartz}) of the form
\begin{equation} \label{fR}
{\cal O} \left( \frac{\Lambda}{\mu} \right) ~f(R)~,
\end{equation}
where $f(R)$ falls off more slowly than the exponential in
(\ref{schwartz}). To obtain the desired result, we must first take
the limit $\mu \rightarrow \infty$ at fixed $\Lambda$ before
taking $R \rightarrow \infty$. Therefore, our results only apply
in the limit of asymptotic density.

Although our result precludes breaking of vector symmetries at
asymptotic density in the case of three {\it exactly} massless
quarks \cite{thermo}, it does not necessarily apply to the case
when the quark masses are allowed to be slightly non-zero. In that
case the results depend on precisely how the limits of zero quark
masses and asymptotic density are taken, as we discuss below.

In \cite{Bedaque:2001je} the authors investigate the effect of
quark masses on the CFL phase. These calculations are done in the
asymptotic limit, and are reliable for sufficiently small quark
masses. When $m_u = m_d \equiv m << m_s$ (unbroken $SU(2)$
isospin, but explicitly broken $SU(3)$), one finds a kaon
condensate. The critical value of $m_s$ at which the condensate
forms is $m_s^* \sim m^{1/3} \Delta_0^{2/3}$, where $\Delta_0$ is
the CFL gap (see, in particular, equation (8) of the first paper).
As kaons transform as a doublet under isospin, the vector $SU(2)$
symmetry is broken in seeming contradiction with our result.

However, a subtle order of limits is at work here. For simplicity,
let us set $m = 0$. Note that the CFL regulator $d$, which was
inserted by hand, explicitly breaks $SU(3)_A$ through color-flavor
locking, leading to small positive mass squared for the pions and
kaons, given as
\begin{eqnarray}
m_{\pi,K}^2 \sim \alpha_s d^2\,\ln\left(\frac{\mu}{d}\right).
\end{eqnarray}
The meson mass is not suppressed by $1/\mu$, since, unlike the
Dirac mass term, the regulator, being a Majorana mass, does not
involve antiquarks~\cite{Hong:1999ei}.

Therefore, even when the light quarks are massless, there is a
critical value of $m_s$ necessary to drive negative the
mass-squared of kaons and cause condensation:
\begin{equation}
m_s^* ~\sim~ \left[ g_s d \mu \, \ln\left(\frac{\mu}{d}\right)
\right]^{1/2} ~>~ ( d \mu )^{1/2}~,
\end{equation}
where $g_s$ is the strong coupling constant. Note the product of
$g_s$ with the logarithm grows as $\mu$ gets large. To obtain our
inequality we must keep the regulator $d$ non-zero until the end
of the calculation in order to see the exponential fall off. To
find the phase with kaon condensation identified
in~\cite{Bedaque:2001je} we must keep $m_s$ larger than $m_s^*$.
(Note $\mu \rightarrow \infty$, so to have any chance of finding
this phase we must take $d \rightarrow 0$ keeping $d R$ large and
$d \mu$ small.)

Since the UV cutoff of the HDET must be larger than $m_s$, we have
\begin{equation}
1 ~>~ \left( \frac{m_s^*}{\Lambda} \right)^2 ~>~ \frac{d}{\Lambda}
\, \frac{\mu}{\Lambda}~,
\end{equation}
which implies
\begin{equation}
\label{fR1}  \frac{\Lambda}{\mu} \, f(R) ~>~ \frac{d}{\Lambda} \,
f(R)~.
\end{equation}
Note the right hand side of this inequality does not necessarily
fall off at large $R$, and also does not go to zero for $\mu
\rightarrow \infty$ at fixed $\Lambda$ and $d$. This is a problem
since to apply our inequality the exponential falloff from
(\ref{schwartz}) must dominate the correction term (\ref{fR}),
which is just the left hand side of (\ref{fR1}). Combining these
equations, we see that the exponential falloff of the correlator
is bounded below,
\begin{equation}
\frac{e^{- 2d \, R}}{d^2} ~>~ {d \over \Lambda}  \, f(R)~,
\end{equation}
in the scaling region with a kaon condensate, $m_s > m_s^*$.

Alternatively, if we had taken $m_s$ to be finite for fixed
regulator $d$ (so that, as $\mu \rightarrow \infty$, eventually
$m_s < m_s^*$), the inequality in (\ref{schwartz}) could be
applied to exclude a Nambu-Goldstone boson, but we would find
ourselves in the phase without a kaon condensate.



\section*{Acknowledgements}
The author wishes to thank Mark Alford, Phillipe de Forcrand,
Simon Hands, Krishna Rajagopal,
Francesco Sannino, Thomas Sch\"afer for useful discussions. The author is thankful
especially to Steve Hsu for the critical discussions and for the
collaboration, upon which some of this lecture is based.
This work  is supported
by KOSEF grant number R01-1999-000-00017-0.

\newpage


\begin{thebibliography}{99}
\bibitem{thermo} To investigate spontaneous symmetry breaking, one
ordinarily has to start at finite volume and insert a source which
explicitly breaks the symmetry. The source is removed only after
the infinite volume limit is taken. We stress that the source does
not have to be a quark mass (it could be a higher dimension
operator), so one can investigate symmetry breaking even when the
quark mass is exactly zero throughout the calculation. (To be
precise, a quark mass does not explicitly violate vector
symmetries, so it cannot play the role of the source in the
thermodynamic limit needed here.)
\end{thebibliography}
\end{document}